\documentclass{iopjournal}

\usepackage[T1]{fontenc}
\usepackage{cite}
\usepackage{amssymb,amsthm,amsmath}

\begin{document}

\articletype{Paper} 

\title{Primordial deuterium abundance from calculations of $p(n,\gamma)$ and $d(p,\gamma)$ reactions within potential-model approach}

\author{Nguyen Le Anh$^{1,*}$\orcid{0000-0002-1198-9921}, Dao Nhut Anh$^1$\orcid{0009-0000-6632-7004}, Hoang Thai An$^1$\orcid{0009-0000-6039-154X}, Nguyen Gia Huy$^1$\orcid{0009-0002-8119-6158} and Bui Minh Loc$^{2}$\orcid{0000-0002-2609-1751}}

\affil{$^1$Department of Physics, Ho Chi Minh City University of Education, 280 An Duong Vuong, Cho Quan Ward, Ho Chi Minh City, Vietnam}

\affil{$^2$San Diego State University, 5500 Campanile Drive, San Diego, CA 92182, USA}

\affil{$^*$Author to whom any correspondence should be addressed.}

\email{anhnl@hcmue.edu.vn}


\begin{abstract}
The $p(n,\gamma)$ and $d(p,\gamma)$ reactions are key nuclear inputs for Big Bang nucleosynthesis. In this work, both reactions are analyzed within a consistent two-body potential framework based on the Malfliet-Tjon interaction, including contributions from both $E1$ and $M1$ transitions. A single scaling factor $\lambda$ controlling the low-energy scattering dynamics is constrained by the $p(n,\gamma)$ and propagated consistently to the $d(p,\gamma)$. The obtained abundance, $\mathrm{D/H} = 2.479^{+0.350}_{-0.177}\times 10^{-5}$, is in good agreement with values inferred from metal-poor damped Lyman-$\alpha$ systems. The modest variations of $\lambda$ lead to a significant change in the predicted $\mathrm{D/H}$ ratio and light-element abundances.
\end{abstract}

\section{Introduction} \label{sec:introduction}

Big Bang nucleosynthesis (BBN) links early-universe physics to the observed abundances of light nuclei, with deuterium providing one of the most precise cosmological probes~\cite{cooke2018,cyburt2016}. Its primordial abundance is highly sensitive to the nuclear reaction rates that govern both its production and destruction. The neutron-proton ($np$) radiative-capture reaction, $p(n,\gamma)$, initiates nucleosynthesis by ending the deuterium bottleneck, while the proton-deuteron ($pd$) radiative-capture reaction, $d(p,\gamma)$, regulates deuterium burning. Accurate cross sections for these two processes are therefore essential for reducing nuclear uncertainties in BBN predictions and for improving constraints on the cosmic baryon density~\cite{serpico2004,cyburt2016,fields2020,pisanti2021}.

Experimental data for the radiative-capture reactions relevant to BBN remain limited at low energies. For $d(p,\gamma)$, the Coulomb barrier suppresses the cross section, and although the Laboratory for Underground Nuclear Astrophysics (LUNA collaboration) provided the most precise measurements to date~\cite{casella2002,mossa2020}, the data remain sparse around $100$~keV~\cite{griffiths1963,warren1963,bailey1970,schmid1995,ma1997,bystritsky2015,tisma2019,turkat2021}, where BBN sensitivity is highest. For $p(n,\gamma)$, only a few low-energy capture measurements exist~\cite{suzuki1995,nagai1997}, and additional photodisintegration data near threshold do not sufficiently constrain the reaction in the BBN window~\cite{schreiber2000,hara2003,moreh1989,tornow2003,chen2026}. As a result, theoretical calculations are essential for supplying accurate reaction rates, especially as increasingly precise primordial deuterium measurements from cosmic microwave background (CMB) analyses and metal-poor systems demand nuclear inputs of matching accuracy.

Methodologically, several modern theoretical approaches have investigated the astrophysical implications of updated $np$ and $pd$ radiative-capture rates~\cite{carlson1998,descouvemont2004,ando2006,johnson2001,rupak2000,beane2015,acharya2022,marcucci2006,marcucci2016,moscoso2021,huang2010,dubovichenko2009,dubovichenko2020}. State-of-the-art calculations based on \textit{ab initio} methods or chiral effective field theory ($\chi$EFT), which are well suited for few-body systems, aim to provide high-precision predictions for low-energy cross sections that serve as critical inputs for BBN. The theoretical uncertainties associated with the two reactions at BBN energies remain markedly different, reflecting the distinct challenges in modeling each process. Among the approaches, the potential model (PM) provides a simple and intuitive framework for describing direct radiative capture~\cite{huang2010}. The PM studies revisited the astrophysical $S$ factor of the $d(p,\gamma)$ reaction, highlighting the important role of the $M1$ transition at very low energies and comparing reaction rates up to $1$~GK with those from other frameworks~\cite{leanh2024}. 

While previous PM studies have focused mainly on reproducing cross sections or astrophysical $S$ factors for individual reactions, their implications for cosmological observables have not been systematically explored. In the present work, the PM approach~\cite{dubovichenko2009,dubovichenko2020,huang2010,leanh2024} is used to describe the $p(n,\gamma)$ and $d(p,\gamma)$ radiative-capture reactions and to supply reliable nuclear inputs for astrophysical applications. Both reactions are treated within a two-body framework based on the Malfliet-Tjon (MT) interaction~\cite{malfliet1969,malfliet1970,awasthi2024}, which offers a practical and robust description of the $A=2,3$ systems. The $E1$ and $M1$ transitions are calculated from bound and scattering wave functions obtained by solving the two-body Schr\"odinger equation in the center-of-mass frame. The bound-state potential is fixed by reproducing the empirical nucleon separation energy, while the scattering potential is generated by scaling the same interaction with a factor $\lambda$. The value of $\lambda$ is constrained by the measured thermal capture cross section of the $p(n,\gamma)$ reaction. The present model is not intended as a microscopic description of nuclear dynamics, but as a simplified framework constrained to low-energy data for astrophysical use.

Motivated by a cluster picture, the effective $pd$ scattering interaction is assumed to arise from the underlying $pp$ and $np$ interactions. At energies around $100$~keV, the deuteron can be treated as a point-like object at leading order. In this limit, the effective $pd$ interaction is approximated by the sum $V_{pd} = V_{np} + V_{pp}$, which is formally equivalent to a folding procedure with a delta-function deuteron density. Within the impulse approximation, the three-body effects entering only at higher order are neglected in the present work.

Rather than quoting a single primordial deuterium abundance, we explore the sensitivity of BBN predictions by examining how the $\mathrm{D/H}$ ratio varies with the scaling factor $\lambda$. For each value of $\lambda$, the calculated cross sections are used to derive Maxwellian-averaged reaction rates, which are parametrized analytically and implemented in the PArthENoPE program~\cite{pisanti2008,consiglio2018,gariazzo2022}. The resulting primordial $\mathrm{D/H}$ values are of the correct order of magnitude and illustrate the dependence of BBN predictions on the low-energy behavior of the $d(p,\gamma)$ reaction. The impact of nuclear-model uncertainties is assessed by constraining $\lambda$ using the experimental uncertainty of the thermal $p(n,\gamma)$ capture cross section and propagating the corresponding variation to the $d(p,\gamma)$ rate and the predicted deuterium abundance.

The remainder of this paper is organized as follows. Section~\ref{sec:method} describes the theoretical framework and numerical implementation. Section~\ref{sec:results} presents the calculated cross sections, astrophysical $S$ factors, reaction rates, and their implications for primordial light-element abundances. Section~\ref{sec:conclusion} summarizes the main conclusions.

\section{Potential-model approach} \label{sec:method}
\subsection{Cross section}

The cross section for the electromagnetic dipole $\Omega 1$ transitions to a bound state is expressed as
\begin{equation}\label{eq:sigmaE}
    \sigma (E) =  \frac{16\pi}{9}  \frac{k_\gamma^3}{\hbar v}\frac{1}{(2S_P+1)(2S_T+1)} \sum_{\Omega} \sum_{\ell_i j_i J_i} |M_{\Omega 1}(E)|^2,
\end{equation}
where the sum over $\Omega$ includes $\Omega \equiv E$ for electric dipole transitions and $\Omega \equiv M$ for magnetic dipole transitions. The photon wave number, $k_\gamma$, is a continuous function of the bombarding energy $E$. The intrinsic spins $S_P$ and $S_T$ correspond to the projectile and target, respectively. The reduced matrix elements for the $\Omega 1$ transition read as
\begin{equation}\label{Mstart}
    M_{\Omega 1} = \langle [S_T \otimes (\ell_f \otimes S_P)_{j_f}]_{J_f} || \mathcal{O}_{\Omega 1} || [S_T \otimes (\ell_i \otimes S_P)_{j_i}]_{J_i}\rangle,
\end{equation}
where the initial scattering state is represented by $| [S_T \otimes (\ell_i \otimes S_P)_{j_i}]_{J_i}\rangle$, and the final bound state by $| [S_T \otimes (\ell_f\otimes S_P)_{j_f}]_{J_f} \rangle$. Here, the total relative angular momentum of the scattering state is $\vec{j}_i = \vec{\ell}_i + \vec{S}_P$ being the relative orbital angular momentum. The scattering channel spin results from the coupling $\vec{J}_i = \vec{S}_T + \vec{j}_i$. For the bound state, the corresponding quantum numbers are obtained by replacing the index $i$ with $f$.

For the $E1$ transition, the operator is given by 
\begin{equation}\label{OElambda}
    \mathcal{O}_{E1} = C_e r Y_1^\mu,
\end{equation}
where $C_{e} = (Z_P/m_P - Z_T/m_T) m e$ is the effective charge, with $m$ being the reduced mass, $e$ the elementary charge, and $Z_P$, $Z_T$, $m_P$, and $m_T$ the charges and masses of projectile and target, respectively. $Y_1^\mu$ ($\mu = -1, 0, 1$) represents the spherical harmonic functions. The matrix element $M_{E1}$ is reduced to the single-particle (s.p.) reduced matrix element as
\begin{align}\label{MElambda2}
    M_{E1} = C_{e} \sqrt{\dfrac{3}{4\pi}}
    (-1)^{S_T+j_f+J_i+1}
    \hat{J}_i\hat{J}_f
    \left\{ \begin{matrix} 
        j_f & J_f & S_T \\ J_i & j_i & 1 
    \end{matrix} \right\}
    M_{0,E1}^{(\mathrm{s.p.})},
\end{align}
where the hat notation is defined as $\hat{J} = \sqrt{2J + 1}$, and $\left\{ \begin{matrix} 
        j_f & J_f & S_T \\ J_i & j_i & 1 
\end{matrix} \right\}$ represents the Wigner $6j$ coefficient.

The operator for the $M1$ transition including orbital and spin components is given by~\cite{leanh2022}
\begin{equation}\label{OM1}
    \mathcal{O}_{M1} = \sqrt{\frac{3}{4\pi}} \left(C_m {\hat{\ell}_\mu} + 2\mu_P {\hat{S}_{P,\mu}} + 2\mu_T {\hat{S}_{T,\mu}}\right),
\end{equation}
where the effective magnetic moment is $C_{m} = (Z_P/m_P^2+Z_T/m_T^2)m \mu_N$, with $\mu_N$ being the nuclear magneton. The magnetic moments of interacting particles used are $\mu_p = 2.793\mu_N$, $\mu_n = -1.913\mu_N$, and $\mu_d = 0.857\mu_N$~\cite{olive2014}. The reduced matrix elements are expressed as \cite{angulo1999,leanh2022}
\begin{equation}
    M_{M1} = \sqrt{\dfrac{3}{4\pi}} 
    \left(M_{M1}^{(0)} + M_{M1}^{(1)} + M_{M1}^{(2)} \right), \label{M1}
\end{equation}
in which
\begin{align}
    M_{M1}^{(0)} &= C_m (-1)^{S_T+j_i+J_f+1} \hat{J}_i\hat{J}_f 
    \left\{\begin{matrix} 
        j_f & J_f & S_T \\ J_i & j_i & 1 
    \end{matrix} \right\}
    M_{0,M1}^{(\mathrm{s.p.})}, \label{MM10}
    \\
    M_{M1}^{(1)} &= \mu_P (-1)^{S_T+j_i+J_f+1} \hat{J}_i\hat{J}_f 
    \left\{\begin{matrix} 
        j_f & J_f & S_T \\ J_i & j_i & 1 
    \end{matrix} \right\}
    M_{1,M1}^{(\mathrm{s.p.})}, \label{MM11}
    \\
    M_{M1}^{(2)} &= \mu_T \delta_{j_ij_f} (-1)^{S_T+j_f+J_i+1} \hat{J}_i\hat{J}_f  
    \left\{\begin{matrix} 
        S_T & J_i & j_f \\ J_f & S_T & 1 
    \end{matrix} \right\} \hat{S}_T\tilde{S}_T
    M_{2,M1}^{(\mathrm{s.p.})}, \label{MM12}
\end{align}
with $\tilde{S}_T=\sqrt{S_T(S_T+1)}$.

The s.p. reduced matrix elements in Eqs.~\eqref{MElambda2}, \eqref{MM10}, \eqref{MM11}, and \eqref{MM12} include the geometrical coefficients and the radial overlap integrals. These matrix elements are expressed as
\begin{align}
    M^{(\mathrm{s.p.})}_{i,\Omega 1} = A_{i,\Omega 1}\cdot I_{\Omega 1},
\end{align}
where the geometrical coefficients $A_{i,\Omega 1}$ ($i=0,1,2$) are
\begin{align} \label{Elam}
    A^{(\mathrm{s.p.})}_{0,E1} &=
    \hat{j}_f \hat{\ell}_f \hat{\ell}_i
    \left( \begin{matrix} 
        \ell_f & 1 & \ell_i  \\ 0 & 0 & 0 
    \end{matrix} \right)
    \left\{ \begin{matrix} 
        \ell_f & j_f & S_P \\ j_i & \ell_i & 1 
    \end{matrix} \right\}, \\
    A^{(\mathrm{s.p.})}_{0,M1} &= \delta_{\ell_i \ell_f} (-1)^{\ell_f + S_P + j_i + 1} \hat{j}_i\hat{j}_f
    \left\{\begin{matrix} 
        \ell_f & j_f & S_P \\ j_i & \ell_i & 1 
    \end{matrix} \right\} \hat{\ell}_i \tilde{\ell}_i, \label{A0sp}
    \\
    A_{1,M1}^{(\mathrm{s.p.})} &= \delta_{\ell_i \ell_f}
    (-1)^{\ell_i + S_P + j_f + 1} \hat{j}_i\hat{j}_f
    \left\{\begin{matrix} 
        S_P & j_f & \ell_i \\ j_i & S_P & 1 
    \end{matrix} \right\}
    \hat{S}_P\tilde{S}_P,
    \label{A1sp}
    \\
    A_{2,M1}^{(\mathrm{s.p.})} & = \delta_{\ell_i \ell_f}\delta_{j_ij_f}, \label{A2core}
\end{align}
with $\left( \begin{matrix} 
        \ell_f & 1 & \ell_i  \\ 0 & 0 & 0 
\end{matrix} \right)$ being $3j$ coefficient.

The radial s.p. overlap integrals for the $E1$ and $M1$ transitions are written as
\begin{align}\label{I1}
    I_{E1} &= \int \phi_{n \ell_f j_f}(r) \chi_{\ell_i j_i}(E,r)r \,dr,\\
    I_{M1} &= \int \phi_{n \ell_f j_f}(r) \chi_{\ell_i j_i}(E,r) \,dr,
\end{align}
where $\chi_{\ell_i j_i}$ and $\phi_{n \ell_f j_f}$ are the scattering and bound wave functions, respectively. These functions are derived from solving the scattering and bound equations with corresponding two-body potentials in the continuous-positive and discrete-negative energy regions, respectively.

\subsection{Interaction potential}
The scattering and bound potentials are expressed as 
\begin{align} 
V(r) = V_{\mathrm{MT}}(r) + V_{\mathrm{C}}(r), 
\end{align}
where $V_{\mathrm{MT}}(r)$ is the Malfliet-Tjon potential~\cite{malfliet1969,malfliet1970}
\begin{align} 
V_{\mathrm{MT}}(r) =  V_{r}\dfrac{e^{-\mu_r r}}{r} - V_{a}\dfrac{e^{-\mu_a r}}{r}, 
\end{align}
with $\mu_r = 2\mu_a$. The depths $V_a$ and $V_r$ are of attractive and repulsive parts, respectively. Spin-orbit interactions are not included in the present work, as the radiative-capture processes at BBN energies are dominated by low-energy $s$- and $p$-wave contributions, for which spin-orbit effects are numerically negligible and cannot be constrained by existing data. Scattering ($s$) and bound ($b$) potentials differ only by a scaling factor $\lambda$, 
\begin{equation}
    V_{\mathrm{MT}}^s(r) = \lambda V_{\mathrm{MT}}^b(r).
\end{equation}
The Coulomb potential, which only appears in the $d(p,\gamma)$ reaction, is modeled as that of a uniformly charged sphere, given by
\begin{align} 
V_{\mathrm{C}}(r) = \begin{cases} \dfrac{Z_P Z_T}{2r_0}\left( 3 - \dfrac{r^2}{r_0^2} \right)e^2, & r < r_0, \\
\dfrac{Z_P Z_T e^2}{r}, & r \geq r_0, \end{cases} 
\end{align}
where $r_0 = 1.25 \times (m_P^{1/3}+m_T^{1/3})$ is in fm.

\subsection{Nuclear reaction rate}
The reaction rates at a given temperature $T$ are~\cite{acharya2025}
\begin{equation} \label{eq:rate}
    R(T)  = \sqrt{\dfrac{8}{m\pi}}N_A(k_BT)^{-3/2} \int_0^\infty g(E)S(E)  \,dE,
\end{equation}
where $N_A$ and $k_B$ are the Avogadro and Boltzmann constants, respectively. The astrophysical $S$ factor is defined as
\begin{align} \label{eq:S_def}
    S(E) = \sigma(E)Ee^{2\pi\eta(E)},
\end{align}
where $\eta(E)$ is the Sommerfeld parameter. 
The Gamow window function is expressed as
\begin{equation} \label{eq:GM}
    g(E)= \exp \left[-\dfrac{E}{k_BT}-2\pi\eta(E)  \right].
\end{equation}

The nuclear reaction rates for the $p(n,\gamma)$ ($\mathrm{png}$) and $d(p,\gamma)$ ($\mathrm{dpg}$) can be approximated analytically~\cite{serpico2004,pisanti2021}
\begin{align} \label{eq:ana_rate}
    R_{\mathrm{png}}(T) &\approx a_0\left(1 + \sum_{i=1}^{6} a_{i} T^{i/2}\right), \\
    R_{\mathrm{dpg}}(T) &\approx T^{-2/3}\exp\left(-\dfrac{a_0}{T^{1/3}}\right) \sum_{i=0}^{17} a_{i+1} T^{i/3},
\end{align}
where $T$ is in GK and $R$ is in cm$^3$ mol$^{-1}$ s$^{-1}$. 

\section{Results and discussion} \label{sec:results}

\subsection{Cross section of $p(n,\gamma)$ reaction}

\begin{table}[]
\setlength{\tabcolsep}{1.2pt}
    \centering
    \caption{Parameters of the bound ($b$) and scattering ($s$) potentials used in the analysis for both reactions. The scaling factor \(\lambda\) relates the nuclear scattering and bound MT potentials through $V_{\mathrm{MT}}^{s}(r)=\lambda V_{\mathrm{MT}}^{b}(r)$.}
    \label{tab:potential}
    \begin{tabular}{ccccccccc} \hline \hline
        Reaction & $r_0$ & $\mu_r$ & $V_r^b$ & $V_r^s$ & $\mu_a$ & $V_a^{b}$ & $V_a^{s}$ & $\lambda$\\
         & [fm] & [fm$^{-1}$] & [MeV] & [MeV] & [fm$^{-1}$] & [MeV] & [MeV] & \\ \hline 
        $p(n,\gamma)$ & $2.506$ & $3.11$ & $1458$ & $1072$ & $1.55$ & $632$ & $464$ & $0.734$ \\
        $d(p,\gamma)$  & $2.831$ & $3.11$ & $1458$ & $2143$ & $1.55$  & $632$ & $929$ & $1.470$ \\ \hline \hline
    \end{tabular}
\end{table}

The cross section of the $p(n,\gamma)$ reaction is obtained by including contributions from both the $E1$ and $M1$ transitions, with the $M1$ component dominating at very low energies. Only transitions to the deuteron ground state are considered. In the present work, the bound state is treated in a pure $s$-wave approximation and is modeled solely by coupled s.p. $1s_{1/2}$ ($\ell_f = 0$) neutron-proton configurations, with the $d$-wave component neglected. Charge-isospin breaking effects are not included explicitly. Instead, their net contribution is effectively absorbed through a very slight adjustment of the depth of the attractive part of the potential to reproduce the experimental binding energy. Because low-energy radiative capture is dominated by the asymptotic tail of the bound-state wave function, such a slight adjustment of the potential depth does not affect the capture matrix elements. The MT potential parameters, $\mu_r = 3.11$~fm$^{-1}$, $V_r^b = 1458$~MeV, $\mu_a = 1.55$~fm$^{-1}$, and $V_a^b = 632$~MeV (see Table~\ref{tab:potential}), yield a deuteron binding energy of $E_b = -2.225$~MeV. These parameters also reproduce the empirical nucleon-nucleon phase shifts reported in Ref.~\cite{stoks1993}. 

The $M1$ transition receives its dominant contribution from incoming $s$ waves ($\ell_i = 0$), while the $E1$ transition proceeds mainly from incoming $p$ waves ($\ell_i = 1$). At the lowest energies, the capture process is therefore dominated by the $M1$ component. It is noted that the $M1$ contribution vanishes if identical potentials are used for the bound and scattering $s$-wave functions, owing to their orthogonality~\cite{blackston2008}. To correctly reproduce the low-energy $np$ scattering properties, the scattering potential is modified by a scaling factor $\lambda \equiv \lambda_{\mathrm{png}} \approx 0.734$, which determines the strengths $V_a^s = 464$~MeV and $V_r^s = 1072$~MeV listed in Table~\ref{tab:potential}. The scaling factor is constrained by the thermal neutron capture cross section at the thermal neutron energy of $E_n = 25.3$~meV, equivalent to $E_{\mathrm{c.m.}} = 1.2625 \times 10^{-8}$~MeV. The calculated thermal cross section is $\sigma_{\mathrm{th}} = 332.6$~mb, in agreement with the experimental value of $332.6 \pm 0.7$~mb~\cite{ando2006}. The small experimental uncertainty ($\delta\sigma_{\mathrm{th}} = 0.7$~mb) translates into a tight constraint on the scaling factor, $\delta\lambda = 5 \times 10^{-5}$. 

\begin{figure}
    \centering
    \includegraphics[width=\linewidth]{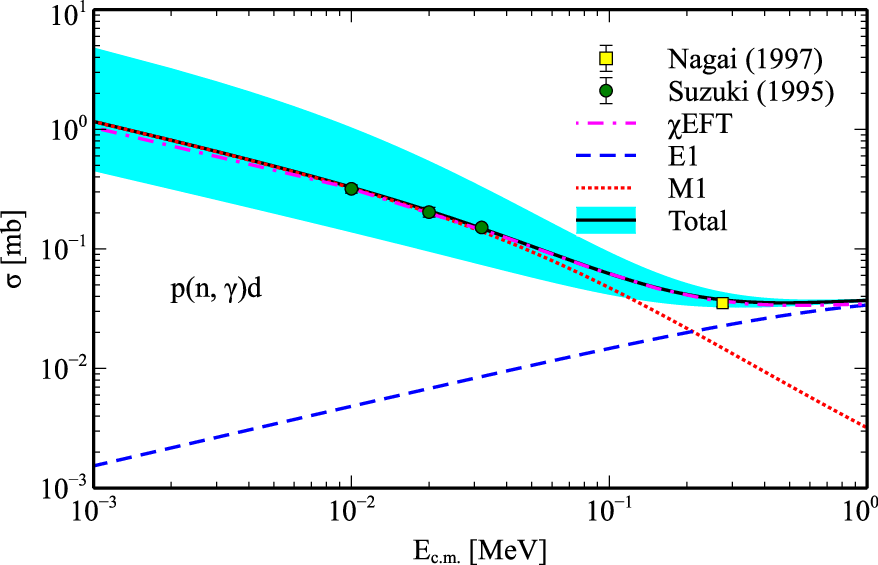}
    \caption{Cross section of $p(n,\gamma)$ reaction below $1$~MeV. The solid line represents the total calculated cross section, while the dashed and dotted lines show the $E1$ and $M1$ contributions, respectively. Experimental data from Suzuki \textit{et al.}~\cite{suzuki1995} (circles) and Nagai \textit{et al.}~\cite{nagai1997} (square) are included for comparison. The dashed-dotted curve shows the $\chi$EFT calculation~\cite{acharya2022}. The shaded band shows the theoretical uncertainty obtained by varying the scaling factor $\lambda = 0.734 \pm 0.023$.}
    \label{fig:png_xs}
\end{figure}

Figure~\ref{fig:png_xs} shows the calculated $p(n,\gamma)$ cross section up to $1$~MeV, together with available experimental data~\cite{suzuki1995,nagai1997} and the $\chi$EFT result~\cite{acharya2022}. Below $100$~keV, the reaction is almost entirely dominated by the $M1$ transition, and the calculated $M1$ contribution reproduces well the low-energy data of Suzuki \textit{et al.}~\cite{suzuki1995}. At higher energies, the $E1$ contribution increases and becomes comparable to the $M1$ component around a few hundred keV.

The data point measured by Nagai \textit{et al.}~\cite{nagai1997} at $E_n = 550$~keV ($E_{\mathrm{c.m.}} = 275$~keV), where the $E1$ and $M1$ components contribute with similar strength, is also well reproduced. That work reported the measured value of $35.2 \pm 2.4~\mu$b. The present calculation yields $37.54~\mu$b, in good agreement with the experimental result, and constrains the scaling factor to $\lambda = 0.745 \pm 0.012$. 

Combining this constraint with the uncertainty extracted from the thermal neutron capture cross section leads to a total model uncertainty of $\delta\lambda = 0.023$, corresponding to $\lambda = 0.734 \pm 0.023$ ($\delta\lambda/\lambda = 0.0313$). This uncertainty band encompasses all available experimental data for the $p(n,\gamma)$ reaction over the energy range considered. As illustrated in Figure~\ref{fig:png_xs}, the cross section at very low energies is highly sensitive to the value of $\lambda$, since in this region the capture process is dominated by the $M1$ transition and is governed by the asymptotic behavior of the scattering wave function. Consequently, small variations in the scaled potential lead to noticeable changes in the overlap between the initial and final states. At higher energies ($E_{\mathrm{c.m.}} > 100$~keV), the $E1$ contribution becomes increasingly important, and the capture process is less sensitive to the low-energy scattering potential, resulting in a significantly reduced dependence on $\lambda$ and a narrower uncertainty band.

It should be emphasized that all allowed $E1$ and $M1$ s.p. transitions satisfying the electromagnetic selection rules are explicitly included in the calculation. The total unpolarized cross section is obtained by summing over the corresponding initial channels leading to the deuteron ground state. A single scaling factor $\lambda$ is employed consistently for both $M1$ and $E1$ contributions, ensuring a unified description of the scattering and bound states. Although experimental studies have investigated the fine splitting of individual $E1$ $p$-wave amplitudes~\cite{blackston2008}, the present analysis focuses on the total cross section relevant for BBN applications and does not introduce channel-dependent renormalization. The total cross section is found to be in good agreement with experimental data over the entire energy range considered, from thermal up to BBN-relevant energies, without the need for channel-dependent or energy-dependent adjustments. This demonstrates that, despite its simplicity, the present model can describe both $M1$-dominated and $E1$-dominated regimes within a consistent framework. 

The physical deuteron contains a small $d$-wave admixture induced by the tensor force, which is included in microscopic and $\chi$EFT calculations. In the present phenomenological framework, the bound state is approximated by its dominant $s$-wave component, consistent with the central nature of the MT interaction~\cite{malfliet1969}. At sub-MeV energies relevant for BBN, radiative capture is governed primarily by low-energy $s$-wave scattering, while higher partial waves are suppressed by the centrifugal barrier. The neglected $d$-wave contribution is therefore expected to have only a minor impact compared to the dominant uncertainty associated with the scattering state.
The obtained cross sections are also in qualitative agreement with the recent $\chi$EFT calculation of Ref.~\cite{acharya2022}, indicating that the dominant low-energy behavior of the $np$ system relevant for radiative capture is reasonably reproduced within this framework.

\subsection{Astrophysical $S$ factor of $d(p,\gamma)$ reaction}

The present calculation for the $d(p,\gamma)$ reaction employs the MT potentials and follows the strategy reported in Ref.~\cite{leanh2024}. Both $E1$ and $M1$ transitions are included, and the $M1$ component provides the dominant contribution at very low energies. The astrophysical $S$ factor is obtained by evaluating the $E1$ and $M1$ matrix elements connecting the relevant scattering states to the $^{3}\mathrm{He}$ ground state. This state has spin-parity $J_f^\pi = 1/2^+$ and is modeled as a proton with spin $S_P = 1/2$ coupled to a deuteron core with intrinsic spin $S_T = 1$.

In the present work, the $A=3$ system is treated as a structureless two-body system composed of a proton and a deuteron cluster. The binding energy of the $1s_{1/2}$ proton state ($\ell_f = 0$) is adjusted to reproduce the experimental reaction $Q$ value, $E_b = -5.493$~MeV. The parameters of the bound-state potential are summarized in Table~\ref{tab:potential}. Remarkably, the same set of nuclear bound-state parameters used for the $np$ system is found to reproduce the bound-state properties of the $pd$ system. The only difference between the two systems arises from the inclusion of the Coulomb interaction in the $pd$ case.

The $E1$ transition corresponds to capture from incoming $p$-wave states with $\ell_i = 1$, while the $M1$ transition corresponds to capture from incoming $s$-wave states with $\ell_i = 0$. The $M1$ transition becomes important at low energies because $s$-wave capture is not suppressed by the centrifugal barrier. As discussed in Ref.~\cite{leanh2024}, the $M1$ strength is very sensitive to the depth of the scattering potential. This reflects the sensitivity of low-energy $s$-wave amplitudes to the behavior of the nuclear interaction.

\begin{figure}
    \centering
    \includegraphics[width=\linewidth]{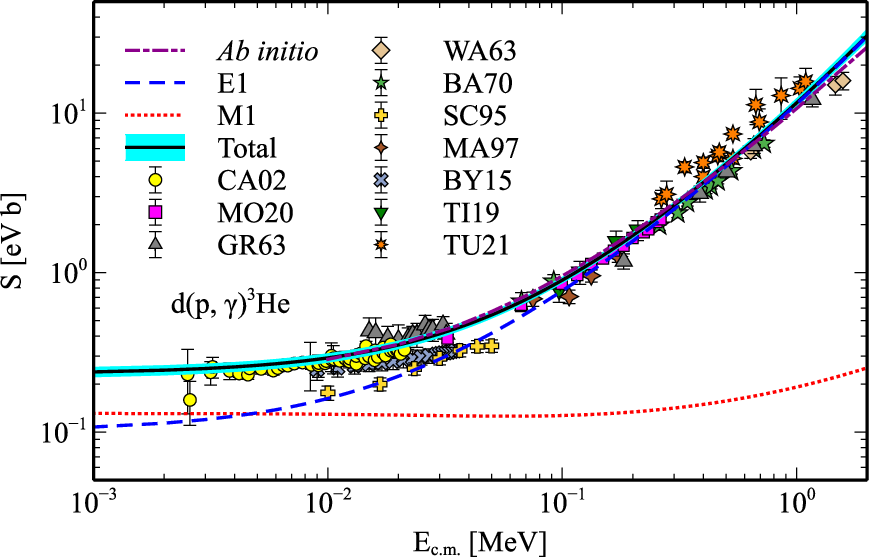}
    \caption{Astrophysical $S$ factor of $d(p,\gamma)$ reaction below $2$~MeV. The results are compared to experimental data~\cite{casella2002,mossa2020,griffiths1963,warren1963,bailey1970,schmid1995,ma1997,bystritsky2015,tisma2019,turkat2021} and the \textit{ab initio} calculation~\cite{marcucci2016}. The scaling factor is $\lambda = 1.470 \pm 0.046$.}
    \label{fig:dpg_sfactor}
\end{figure}

Figure~\ref{fig:dpg_sfactor} presents the calculated astrophysical $S$ factor for the $d(p,\gamma)$ reaction. The low-energy scattering interaction in the $d(p,\gamma)$ is taken to be governed by the same underlying dynamics as in the $p(n,\gamma)$ system. Accordingly, the MT scattering potential is scaled by a factor
$\lambda \equiv \lambda_{\mathrm{dpg}} = 2\lambda_{\mathrm{png}} \approx 1.470 \pm 0.046$ due to $V_{pd} = V_{np} + V_{pp} = 2V_{np}$ and used consistently for both the $E1$ and $M1$ transitions. The calculated results are in good agreement with the available experimental data, in particular with the high-precision measurements of Refs.~\cite{casella2002,mossa2020}. With the potential strengths $V_a^s = 929$~MeV and $V_r^s = 2143$~MeV corresponding to $\lambda = 1.470$, the calculated $S(0)$ is in good agreement with the expected value without spectroscopic factor. The value of $S(0) = 0.228 \pm 0.015$~eV\,b exceeds recent determinations by about $4\%$~\cite{turkat2021,moscoso2021}, while remaining consistent within uncertainties. The total $S$ factor obtained below $500$ keV is represented by the cubic approximation $S(E) \approx 0.228 + 5.946E + 7.929E^2-3.145E^3$, where $S$ and $E$ are given in eV\,b and MeV, respectively. 

In addition, the \textit{ab initio} results in Ref.~\cite{marcucci2016} lie systematically above the present calculation by approximately $5\%$. This difference can be attributable to the more complete treatment of many-body currents and nuclear correlations in the \textit{ab initio} framework. The shaded band in Figure~\ref{fig:dpg_sfactor} represents the theoretical uncertainty associated with the scaling factor, $\delta\lambda \approx 0.046$. As shown, variations of $\lambda$ within this range lead to only a modest change in the astrophysical $S$ factor over the energy region of interest. The resulting uncertainty band is sufficient to encompass the available experimental data in LUNA measurements~\cite{casella2002,mossa2020}.

\subsection{Reaction rates and light-element abundances}

\begin{table}[]
    \centering
    \caption{Predicted reaction rates of $p(n,\gamma)$ and $d(p,\gamma)$ reactions below $4$~GK. The temperatures and reaction rates are given in GK and cm$^3$\,mol$^{-1}$\,s$^{-1}$, respectively.}
    \label{tab:RR}
    \begin{tabular}{cccccc} \hline \hline
        $T$ & $R_{\text{png}}$ & $R_{\text{dpg}}$ & $T$ & $R_{\text{png}}$ & $R_{\text{dpg}}$ \\ 
        {}[GK] & [cm$^3$\,mol$^{-1}$\,s$^{-1}$] & [cm$^3$\,mol$^{-1}$\,s$^{-1}$] & [GK] & [cm$^3$\,mol$^{-1}$\,s$^{-1}$] & [cm$^3$\,mol$^{-1}$\,s$^{-1}$] \\ \hline
        $0.001$ & $4.34 \times 10^4$ & $1.51 \times 10^{-11}$ & $0.11$ & $3.76 \times 10^{4}$ & $7.51 \times 10^{0}$ \\
        $0.002$ & $4.34 \times 10^4$ & $2.08 \times 10^{-8}$ & $0.12$ & $3.72 \times 10^4$ & $9.16 \times 10^{0}$ \\
        $0.003$ & $4.34 \times 10^4$ & $6.76 \times 10^{-7}$ & $0.13$ & $3.68 \times 10^4$ & $1.09 \times 10^{1}$ \\
        $0.004$ & $4.34 \times 10^4$ & $5.99 \times 10^{-6}$ & $0.14$ & $3.64 \times 10^4$ & $1.29 \times 10^{1}$ \\
        $0.005$ & $4.34 \times 10^4$ & $2.80 \times 10^{-5}$ & $0.15$ & $3.60 \times 10^4$ & $1.49 \times 10^{1}$ \\
        $0.006$ & $4.34 \times 10^4$ & $9.07 \times 10^{-5}$ & $0.16$ & $3.57 \times 10^4$ & $1.70 \times 10^{1}$ \\
        $0.007$ & $4.34 \times 10^4$ & $2.31 \times 10^{-4}$ & $0.18$ & $3.50 \times 10^4$ & $2.17 \times 10^{1}$ \\
        $0.008$ & $4.33 \times 10^4$ & $4.98 \times 10^{-4}$ & $0.20$ & $3.44 \times 10^4$ & $2.00 \times 10^{1}$ \\
        $0.009$ & $4.33 \times 10^4$ & $9.52 \times 10^{-4}$ & $0.25$ & $3.30 \times 10^4$ & $4.01 \times 10^{1}$ \\
        $0.010$ & $4.32 \times 10^4$ & $1.66 \times 10^{-3}$ & $0.30$ & $3.19 \times 10^4$ & $5.70 \times 10^{1}$ \\
        $0.011$ & $4.31 \times 10^4$ & $2.70 \times 10^{-3}$ & $0.35$ & $3.10 \times 10^4$ & $7.46 \times 10^{1}$ \\
        $0.012$ & $4.31 \times 10^4$ & $4.15 \times 10^{-3}$ & $0.40$ & $3.01 \times 10^4$ & $9.36 \times 10^{1}$ \\
        $0.013$ & $4.30 \times 10^4$ & $6.09 \times 10^{-3}$ & $0.45$ & $2.94 \times 10^4$ & $1.14 \times 10^{2}$ \\
        $0.014$ & $4.29 \times 10^4$ & $8.60 \times 10^{-3}$ & $0.50$ & $2.88 \times 10^4$ & $1.35 \times 10^{2}$ \\
        $0.015$ & $4.29 \times 10^4$ & $1.18 \times 10^{-2}$ & $0.60$ & $2.79 \times 10^4$ & $1.80 \times 10^{2}$ \\
        $0.016$ & $4.28 \times 10^4$ & $1.57 \times 10^{-2}$ & $0.70$ & $2.71 \times 10^4$ & $2.29 \times 10^{2}$ \\
        $0.018$ & $4.27 \times 10^4$ & $2.60 \times 10^{-2}$ & $0.80$ & $2.66 \times 10^4$ & $2.79 \times 10^{2}$ \\
        $0.020$ & $4.25 \times 10^4$ & $4.01 \times 10^{-2}$ & $0.90$ & $2.62 \times 10^4$ & $3.32 \times 10^{2}$ \\
        $0.025$ & $4.22 \times 10^4$ & $9.57 \times 10^{-2}$ & $1.00$ & $2.59 \times 10^4$ & $3.88 \times 10^{2}$ \\
        $0.030$ & $4.19 \times 10^4$ & $1.85 \times 10^{-1}$ & $1.25$ & $2.55 \times 10^4$ & $5.32 \times 10^{2}$ \\
        $0.040$ & $4.12 \times 10^4$ & $4.85 \times 10^{-1}$ & $1.50$ & $2.55 \times 10^4$ & $6.85 \times 10^{2}$ \\
        $0.050$ & $4.06 \times 10^4$ & $9.62 \times 10^{-1}$ & $1.75$ & $2.57 \times 10^4$ & $8.45 \times 10^{2}$ \\
        $0.060$ & $4.00 \times 10^4$ & $1.62 \times 10^{0}$ & $2.00$ & $2.61 \times 10^4$ & $1.01 \times 10^{3}$ \\
        $0.070$ & $3.95 \times 10^4$ & $2.46 \times 10^{0}$ & $2.50$ & $2.71 \times 10^4$ & $1.35 \times 10^{3}$ \\
        $0.080$ & $3.90 \times 10^4$ & $3.48 \times 10^{0}$ & $3.00$ & $2.83 \times 10^4$ & $1.70 \times 10^{3}$ \\
        $0.090$ & $3.85 \times 10^4$ & $4.67 \times 10^{0}$ & $3.50$ & $2.95 \times 10^4$ & $2.04 \times 10^{3}$ \\
        $0.100$ & $3.80 \times 10^4$ & $6.01 \times 10^{0}$ & $4.00$ & $3.06 \times 10^4$ & $2.35 \times 10^{3}$ \\
        \hline \hline  
    \end{tabular}
\end{table}

The reaction rates for both $p(n,\gamma)$ and $d(p,\gamma)$ are calculated up to $T=10$~GK, with integration performed over energies up to $2$~MeV. The numerical values for both reactions from $0.001$~GK to $4$~GK are listed in Table~\ref{tab:RR}. In our calculation, the reaction rates for both reactions exhibit only weak temperature dependence above $8$~GK. The reaction rates are therefore assumed to remain constant for $T>10$~GK, with the values at $T=10$~GK, $3.17 \times 10^{4}$~cm$^{3}$\,mol$^{-1}$\,s$^{-1}$ for $p(n,\gamma)$ and $3.73 \times 10^{3}$~cm$^{3}$\,mol$^{-1}$\,s$^{-1}$ for $d(p,\gamma)$, adopted at higher temperatures.

\begin{table}[]
    \centering
    \caption{Coefficients $a_i$ used in the analytical reaction-rate parametrizations 
    $R_{\mathrm{png}}(T)$ and $R_{\mathrm{dpg}}(T)$. 
    The fits are valid up to $T = 10$~GK.}
    \label{tab:ai}
    \begin{tabular}{crcr} 
    \hline \hline
    \multicolumn{4}{c}{$p(n,\gamma)$} \\ \hline 
    $i$ & $a_i$ & $i$ & $a_i$ \\ \hline 
    $0$ & $46384.1755$ & $4$ & $-0.3858$ \\
    $1$ & $-0.5396$ & $5$ & $0.0778$ \\
    $2$ & $-0.3765$ & $6$ & $-0.0058$ \\
    $3$ & $0.7870$ &  &  \\ \hline
    \multicolumn{4}{c}{$d(p,\gamma)$} \\ \hline 
    $i$ & $a_i$ & $i$ & $a_i$ \\ \hline 
    $0$ & $4.9476$ & $10$ & $-84.6097$ \\
    $1$ & $-209.4585$ & $11$ & $-37687.4043$ \\
    $2$ & $497230.8185$ & $12$ & $66.8863$ \\
    $3$ & $-1460914.9799$ & $13$ & $51.1938$ \\
    $4$ & $1448197.8718$ & $14$ & $8614.1906$ \\
    $5$ & $-625.3102$ & $15$ & $-41.0109$ \\
    $6$ & $-620950.9505$ & $16$ & $-3805.1857$ \\
    $7$ & $517.9641$ & $17$ & $1574.4794$ \\
    $8$ & $222848.7887$ & $18$ & $-193.3532$ \\
    $9$ & $-0.0000$ &  & \\
    \hline \hline 
    \end{tabular}
\end{table}

The coefficients $a_i$ used to parametrize the analytical reaction rates $R_{\mathrm{png}}(T)$ and $R_{\mathrm{dpg}}(T)$ are given in Table~\ref{tab:ai}. These parametrizations are valid up to $10$~GK for use in BBN calculations. Assuming a relative uncertainty of $5\%$ for the tabulated rates, the quality of the fits is quantified by reduced chi-squared values of $\chi^2_\nu = 1.244 \times 10^{-3}$ for $R_{\mathrm{png}}(T)$ and $\chi^2_\nu = 6.719 \times 10^{-1}$ for $R_{\mathrm{dpg}}(T)$ over the temperature range from $0.01$~GK to $10$~GK.

\begin{table}[]
    \centering
    \caption{Primordial $\mathrm{D/H}$ abundance ratio from BBN prediction and observation. The uncertainty in this work is from the propagated variation of the scaling factor $\lambda$ constrained by the $p(n,\gamma)$ capture cross section ($\delta\lambda/\lambda = 0.0313$).}
    \label{tab:DH}
    \begin{tabular}{lc} 
    \hline\hline
        References & $\mathrm{D/H}$ ($\times 10^{-5}$) \\ \hline
        Cooke \textit{et al.} (2018)~\cite{cooke2018} & $2.527 \pm 0.030$ \\
        Mossa \textit{et al.} (2020)~\cite{mossa2020} & $2.52 \pm 0.09$ \\
        Pitrou \textit{et al.} (2021)~\cite{pitrou2021} & $2.439 \pm 0.037$ \\
        Pisanti \textit{et al.} (2021)~\cite{pisanti2021} & $2.51 \pm 0.07$ \\
        Yeh \textit{et al.} (2021)~\cite{yeh2021} & $2.51 \pm 0.11$  \\
        Shen \& He (2024)~\cite{shen2024} & $2.471 \pm 0.039$  \\
        \hline
        This work & $2.479^{+0.350}_{-0.177}$  \\ \hline \hline
    \end{tabular}
\end{table}

Using the latest public version of PArthENoPE~\cite{gariazzo2022}, the primordial deuterium abundance is calculated with a reaction network including $40$ reactions and $9$ nuclides. The neutron lifetime $\tau_n = 879.4$~s~\cite{zyla2020}, the baryon-to-photon ratio $\eta_{10} = 6.13832$, and the baryon density parameter $\omega_b h^2 = 0.02242$ inferred from the final Planck CMB analysis are adopted~\cite{aghanim2020}. The uncertainty is obtained by propagating the allowed variation of the scaling factor $\lambda$, constrained by the experimental uncertainty of the thermal $p(n,\gamma)$ capture cross section, through the radiative-capture reaction rates.

Because the capture cross sections are obtained by solving the Schr\"odinger equation with a scaled potential, the resulting scattering wave functions depend nonlinearly on the strength parameter $\lambda$. Consequently, the calculated cross section $\sigma(E)$ is not a linear function of $\lambda$, and equal positive and negative variations of $\lambda$ do not produce symmetric changes in the reaction rates. 
Additionally, the thermonuclear reaction rates enter the BBN calculation through a coupled system of differential equations, in which the predicted light-element abundances depend nonlinearly on these rates, particularly on the $d(p,\gamma)$ reaction in the relevant temperature range. This nonlinear propagation yields an asymmetric abundance, $\mathrm{D/H} = 2.479^{+0.350}_{-0.177} \times 10^{-5}$.

Although the relative uncertainty of the scaling factor is modest, $\delta\lambda/\lambda \approx 3\%$, it propagates nonlinearly through the reaction network and leads to a significantly larger variation of the predicted primordial abundance, reaching up to $14\%$. This enlarged uncertainty is primarily attributed to the description of the data point in Ref.~\cite{nagai1997}. If only the thermal-energy constraint is used, corresponding to $\delta\lambda = 5 \times 10^{-5}$, the uncertainty in the primordial abundance is significantly reduced to $\mathrm{D/H} = (2.479 \pm 0.001) \times 10^{-5}$.

Observationally, the deuterium abundance is determined from high-redshift absorption features in metal-poor damped Lyman-$\alpha$ systems. The predicted abundance obtained with the PM is in good agreement with the observational value $(2.527 \pm 0.030) \times 10^{-5}$ in Ref.~\cite{cooke2018}. Table~\ref{tab:DH} compares the BBN predictions with the observational constraints. Our result is very close to the recent BBN prediction of Shen and He, $(2.471 \pm 0.039)\times10^{-5}$~\cite{shen2024}, obtained using the PRIMAT Monte Carlo framework.

\begin{figure}[]
    \centering
    \includegraphics[width=\linewidth]{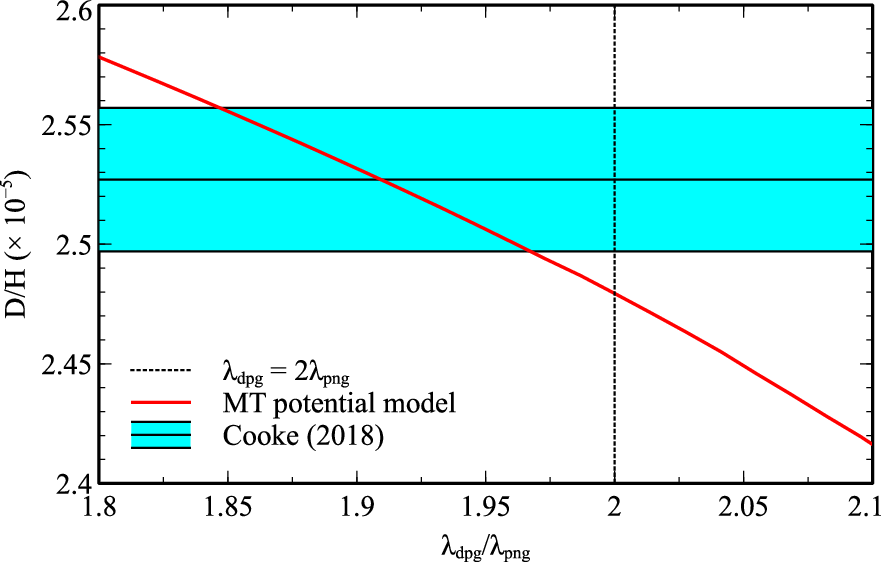}
    \caption{Primordial deuterium abundance $\mathrm{D}/\mathrm{H}$ as a function of the ratio $\lambda_{\mathrm{dpg}}/\lambda_{\mathrm{png}}$ obtained in the present work. The shaded band indicates the observational constraint from Cooke \textit{et al.}~\cite{cooke2018}, while the vertical dashed line marks the central value $\lambda_{\mathrm{dpg}}/\lambda_{\mathrm{png}} = 2$.}

    \label{fig:dpg_lambda_DH}
\end{figure}

Figure~\ref{fig:dpg_lambda_DH} shows the dependence of $\mathrm{D}/\mathrm{H}$ on the ratio $\lambda_{\mathrm{dpg}}/\lambda_{\mathrm{png}}$. The monotonic decrease of $\mathrm{D}/\mathrm{H}$ with increasing $\lambda_{\mathrm{dpg}}/\lambda_{\mathrm{png}}$ reflects enhanced deuterium destruction through the $d(p,\gamma)$ reaction. The overlap with the observational band indicates that the scattering PM renormalization is compatible with observational constraints. The strong sensitivity of $\mathrm{D}/\mathrm{H}$ to the scaling factor underscores the importance of the low-energy renormalization of the $pd$ interaction. Approximating $V_{pd} = 2V_{np}$ is therefore a reasonable and well-motivated approximation. The observed deuterium abundance is reproduced for $\lambda_{\mathrm{dpg}}/\lambda_{\mathrm{png}} = 1.91 \pm 0.06$, which differs from the nominal expectation of $2$ by about $5\%$ and lies within the intrinsic uncertainty of the present model. Moreover, the smooth dependence of $\mathrm{D}/\mathrm{H}$ on $\lambda$ demonstrates the stability of the fitting procedure and supports the validity of the analytical
$d(p,\gamma)$ reaction-rate parametrization adopted for the BBN calculations.

\begin{table}[]
    \centering
    \caption{Comparison of primordial light-element abundances obtained using different radiative-capture rate inputs, with all other nuclear rates taken from Ref.~\cite{pisanti2021}. Our predictions are compared with theoretical results from dEFT~\cite{ando2006} and the $R$-matrix approach~\cite{johnson2001}. Observational constraints are adopted from Ref.~\cite{pitrou2018}.
}
    \label{tab:hydrogen}
    \begin{tabular}{lrrrr} 
    \hline\hline
        Abundances & Observation & Ref.~\cite{ando2006} & Ref.~\cite{johnson2001} & Potential model \\
    \hline
    $Y_p$ & $0.2449$ & $0.24852$ & $0.24849$ & $0.24683^{+0.0005}_{-0.0007}$ \\ 
    $\mathrm{D/H}$ ($\times 10^{-5}$) & $2.527$ & $2.5467$ & $2.5580$ & $2.4794^{+0.3500}_{-0.1773}$ \\
    $^3\mathrm{He}/\mathrm{H}$ ($\times 10^{-6}$) & $<11$ & $10.0921$ & $10.0864$ & $10.4947^{+0.3894}_{-0.6199}$ \\
    $^7\mathrm{Li}/\mathrm{H}$ ($\times 10^{-10}$) & $1.58$ & $4.4646$ & $4.3632$ & $4.7217^{+4.4313}_{-2.8554}$ \\
    \hline \hline
    \end{tabular}
\end{table}

Finally, the obtained light-element abundances are summarized in Table~\ref{tab:hydrogen} and compared with other approaches such as dibaryon effective field theory (dEFT)~\cite{ando2006} and the $R$-matrix method~\cite{johnson2001}. The present results show good agreement with abundances inferred from observational data~\cite{pitrou2018}. The exception is the lithium abundance, which reflects the well-known cosmological lithium problem~\cite{fields2011}. Allowing the scaling factor $\lambda$ to vary within its constrained range leads to a pronounced change in the predicted $^{7}$Li/H abundance, yielding 
$^{7}\mathrm{Li}/\mathrm{H} = 4.72^{+4.43}_{-2.86} \times 10^{-10}$. 
This demonstrates a strong sensitivity of the lithium prediction to variations of the low-energy radiative-capture rates through the propagation in the BBN network. However, even the lowest value obtained within the allowed range remains significantly higher than the observed abundance of $1.58 \times 10^{-10}$. Modifications of the $p(n,\gamma)$ and $d(p,\gamma)$ reaction rates alone are therefore insufficient to resolve the lithium discrepancy within the present framework. The light-element abundances depend on both the $p(n,\gamma)$ and $d(p,\gamma)$ rates, with the latter having the dominant impact on the predicted $\mathrm{D}/\mathrm{H}$ ratio. This is consistent with earlier studies~\cite{ando2006,cyburt2008}, which identified $d(p,\gamma)$ and $^3\mathrm{He}(\alpha,\gamma)$ as dominant sources of theoretical error in BBN predictions. It is also noted that the MT potential previously was applied to studies of the $^{3}\mathrm{He}+\alpha$ system~\cite{awasthi2025}. In addition to uncertainties in nuclear reaction rates, modifications of the particle distributions in the primordial plasma may also influence the predicted abundances within the standard cosmological scenario~\cite{voronchev2026}. Overall, the agreement with observationally inferred abundances supports the validity of the PM approach used to describe the $d(p,\gamma)$ reaction~\cite{dubovichenko2009,dubovichenko2020,huang2010,leanh2024}, which yields reliable $S$ factors and reaction rates for BBN calculations. There is no additional adjustment performed in the $d(p,\gamma)$ once $\lambda$ is fixed by the thermal $p(n,\gamma)$ cross section. Astrophysical observables thus offer a complementary probe of nuclear reaction models at low energies.

\section{Conclusions}
\label{sec:conclusion}

We present a two-body PM description of the $p(n,\gamma)$ and $d(p,\gamma)$ radiative-capture reactions relevant for BBN, based on the MT interaction. The low-energy renormalization of the scattering potential is governed by a single scaling factor constrained by the thermal $p(n,\gamma)$ cross section and propagated consistently to the $d(p,\gamma)$. This approach demonstrates that both reactions can be consistently described by constructing the effective $pd$ scattering interaction from the underlying $np$ subsystem, without requiring independent nuclear inputs.

While variations of the scaling factor induce only modest changes in the $d(p,\gamma)$ astrophysical $S$ factor in the BBN energy range, they give rise to a pronounced sensitivity of the primordial deuterium abundance, underscoring the critical role of low-energy radiative-capture rates in BBN calculations. Implementing the resulting reaction rates in PArthENoPE yields a primordial deuterium abundance of $\mathrm{D}/\mathrm{H} = 2.479^{+0.350}_{-0.177} \times 10^{-5}$, consistent with observational determinations and recent BBN studies. The more precise determination of the low-energy $p(n,\gamma)$ cross section would further tighten the constraint on the scaling factor $\lambda$ and reduce the uncertainty in the predicted primordial deuterium abundance. 

More generally, the present results highlight the importance of accurately describing the first reactions in the BBN network. Variations in the $p(n,\gamma)$ and $d(p,\gamma)$ rates propagate through the reaction chain and can significantly affect the predicted light-element abundances, emphasizing the connection between low-energy nuclear physics and cosmological observations.
An important extension of the present study would be the implementation of a more microscopic nucleon-nucleon interaction capable of describing the $p(n,\gamma)$ and $d(p,\gamma)$ reactions simultaneously within a unified framework. Such an approach would allow for a more direct connection between few-body nuclear dynamics and cosmological observables while preserving the correlated treatment of nuclear uncertainties emphasized in this work.

\ack{This work was funded by Ho Chi Minh City University of Education Foundation for Science and Technology under grant number CS.2025.19.10T{\DJ}. BML is supported by the U.S. Department of Energy, under Award Number DE-NA0004075.}



\data{The data cannot be made publicly available upon publication because no suitable repository exists for hosting data in this field of study. The data that support the findings of this study are available upon reasonable request from the authors.}


\bibliographystyle{iopart-num}
\bibliography{refs}

\end{document}